\documentclass[superscriptaddress,twocolumn,secnumarabic,amsmath,amssymb,nofootinbib,aps,prl]{revtex4-1}
\usepackage{graphicx,color}
\usepackage{bm}%
\usepackage[normalem]{ulem}%
\usepackage{gensymb,bbold}%
\usepackage{datetime}%
\newcommand{\new}{\textcolor{black}}

\newcommand{\qq}{{\bf q}}

\newcommand{\rr}{{\bf r}}

\newcommand{\EE}{{\bf E}}
\newcommand{\Er}{\widetilde{\bf E}}

\newcommand{\Aa}{{\bf A}}

\newcommand{\HH}{{\bf H}}

\newcommand{\Hr}{\widetilde{\bf H}}

\newcommand{\LL}{{\cal L}}

\newcommand{\PP}{{\cal P}}

\newcommand{\be}{\begin{equation}}
\newcommand{\ee}{\end{equation}}
\newcommand{\ba}{\begin{eqnarray}}
\newcommand{\ea}{\end{eqnarray}}
\newcommand{\bse}{\begin{subequations}}
\newcommand{\ese}{\end{subequations}}
\newcommand{\beq}{\begin{eqnarray}}
\newcommand{\eeq}{\end{eqnarray}}

\newcommand{\im}{\mbox{Im}}
\newcommand{\re}{\mbox{Re}}

\begin{document}

\title{The electric and magnetic disordered Maxwell equations as eigenvalue problem}

\author{Walter Schirmacher}
\affiliation{%
Center for Life Nano science@Sapienza, Istituto Italiano di Tecnologia,
Viale Regina Elena, 291, I-00161 Roma, Italy}
\affiliation{%
Institut f\"ur Physik, Universit\"at Mainz, Staudinger Weg 7,
D-55099 Mainz, Germany}
\author{Thomas Franosch}
\affiliation{%
	\mbox{Institut f\"ur Theoretische Physik, Universit\"at Innsbruck,
Technikerstrasse 21A, A-6020 Innsbruck, Austria}}
\author{Marco Leonetti}
\affiliation{%
Center for Life Nano science@Sapienza, Istituto Italiano di Tecnologia,
Viale Regina Elena, 291, I-00161 Roma, Italy}
\affiliation{%
	Soft and Living Matter Laboratory, Institute of Nanotechnology, Consiglio Nazionale delle Ricerche, 00185 Rome, Italy}
\author{Giancarlo Ruocco}
\affiliation{%
Center for Life Nano science@Sapienza, Istituto Italiano di Tecnologia,
Viale Regina Elena, 291, I-00161 Roma, Italy}
\affiliation{Dipartimento di Fisica, Universit\`a di Roma
``La Sapienza'', P'le Aldo Moro 2, I-00185, Roma, Italy}

\begin{abstract}
We consider Maxwell's equations in a 3-dimensional material, in which both, the electric permittivity, as well as the magnetic permeability, fluctuate in space. Differently from all previous treatments of the \textit{disordered} electromagnetic problem, we transform Maxwell's equations and the
	electric and magnetic fields in such a way that the linear operator in the resulting secular equations is manifestly Hermitian, in order to deal with a proper eigenvalue problem. As an application of our general formalism, we use an appropriate version of the Coherent-Potential approximation (CPA) to calculate the photon density of states and scattering-mean-free path. 
	Applying standard localization theory,
	we find that in the presence of both electric and magnetic disorder the spectral range of Anderson localization appears to be much larger than in the case of electric (or magnetic) disorder only. Our result could explain the absence of experimental evidence of 3D Anderson localization of light (all the existing experiments has been performed with electric disorder only) and pave the way towards a successful search of this, up to now, elusive phenomenon.
\pacs{65.60.+a}
\end{abstract}
\maketitle

\section*{Introduction}

Understanding the propagation and scattering of electromagnetic
radiation in random media, especially visible light,
is an issue, which is important in different parts of
science
\cite{chandra60,vandehulst80,ishimaru78,lagendijk96,tregoures02,sheng06,battaglia10}. A particularly
interesting feature of waves in a disordered environment is
the possibility of localization, i.e. the absence of diffusion, demonstrated
first for electron wave functions by Anderson \cite{and58}.
Anderson localization (AL) arises from the interference of the waves scattered
by the random inhomogeneities of the medium 
\cite{abrahams79,lr85,sheng06,evers08,wolfle10}. This phenomenon
occurs with all kinds of waves, including
atomic-matter and gravitational waves 
\cite{billy08,roati08,rothstein13}.

Localization of classical waves has first been discussed
by John et al. \cite{john83,john83a} for acoustical and later for
electromagnetic waves (light) \cite{john84,john87}. The successful
observation of weak localization of light 
(the back-scattering cone) \cite{wolf85}
created
an impact for looking for strong AL of light
\cite{anderson85,wiersma97,sheng06a,lagendijk09,genack11}.
It was realized \cite{abdullaev80,deraedt89} that the chances for the 
observation of this phenomenon are much higher in dimensionally reduced
systems. This has been successfully demonstrated in paraxial structures
with transverse (2-dimensional) disorder \cite{schwartz07,karbasi12}
and two-dimensional photonic crystals
\cite{riboli14}. In 3-dimensional media with a spatially fluctuating
permittivity, however, until now, AL has not been found
\cite{wiersma97,wiersma97,storzer06,sperling12,scheffold13,sperling16,skipetrov16}.
Indeed, 3D localization effects are often 
 obscured by absorption or fluorescence processes, making its experimental demonstration
extremely elusive \cite{skipetrov16}. 
Recently,
the possibility of obtaining Anderson localization in 3D systems
has been made plausible in numerical simulations of
($i$) hyperuniform
amorphous photonic materials
\cite{haberko20,scheffold22}, and ($ii$)
systems with overlapping spherical, perfectly conducting
obstacles
\cite{yamilov22}.

On the other hand, the theoretical description of AL of light
is, until now, built on the ground
of a mathematically questionable mapping of Maxwell's equations to
Anderson's Schr\"odinger equation of an electron in a random potential
\cite{john84,john87}. 
This mapping,
which was taken over by
the subsequent literature
\cite{deraedt89,albada91,baraba92,kroha93,sheng06},
started with the Helmholtz equation for a stationary
frequency-dependent electric field $\EE(\rr,\omega)$
in the presence of a spatially fluctuating permittivity
$\epsilon(\rr)$,
derived from
Maxwell's equations\footnote{%
Here 
$\mu_0=1/\epsilon_0c_0^2$ is the magnetic permeability of the vacuum,
$\epsilon_0$ is the electric permittivity of the vacuum and
$c_0$ is the vacuum light velocity.}
\be\label{helm1}
\omega^2\epsilon(\rr)\,\EE(\rr,\omega)=\frac{1}{\mu_0}\nabla\times[\nabla\times \EE(\rr,\omega)]\, ,
\ee
This equation was transformed in the following way:
\mbox{($i$) the double curl was converted to} $-\nabla^2$, ignoring that
\mbox{$\nabla\cdot\EE\neq 0$}
for $\nabla\epsilon\neq 0$,
($ii$) the coefficient of $\EE$ on the LHS, which features
the spectral parameter $\omega^2$ of the eigenvalue equation, 
was rewritten as $\omega^2\epsilon_0
+\omega^2[\epsilon(\rr)-\epsilon_0]$, where the second term was
re-interpreted as an $\omega$ dependent potential (changing completely
the physical content),
and ($iii$) the eigenvalue problem associated with Eq. (\ref{helm1})
was neither formulated nor solved properly (see below).

In the case of transverse localization this truncated
and ill-posed equation (called ``potential-type approach'' by
Schirmacher {\it et al.}
\cite{schirm18})
produced results, which were at variance with
experiment: A wavelength dependence of the localization length,
predicted on base of this equation \cite{deraedt89,karbasi12a},
was
not observed experimentally and is not predicted by
a proper treatment
\cite{schirm18}.
In view of all these inconsistencies it is apparent that
a mean-field theory, based on a consistently formulated Hermitian eigenvalue
problem of Maxwell theory in the presence of disorder, is urgently
called for.

Here, we present such a mean-field theory of disorder, 
based on a properly formulated eigenvalue problem.
In this theory we allow both for
electric 
(spatially varying electric permittivity $\epsilon(\rr)$)
and magnetic disorder
(spatially varying magnetic permeability $\mu(\rr)$).
The theory is an appropriate version of the coherent-potential
approximation (CPA), derived by S. K\"ohler and two of the present authors
for elastic waves in the presence of disorder \cite{kohler13}.

Applying the CPA results for the
scattering mean-free path and the density of states
to standard localization theory
suggests that by combining electric and magnetic disorder
the chances for observing AL of light in three dimensions
are greatly enhanced with respect to the case where only one quantity ($\epsilon(\rr)$ {\it or}  $\mu(\rr)$) is left to vary.

We start by defining dimensionless electric and magnetic moduli
$M_\epsilon(\rr) := \epsilon_0/\epsilon(\rr)$ and $M_\mu(\rr) := \mu_0/\mu(\rr)$. The generalization of (\ref{helm1}) for including magnetic
disorder takes the form\footnote{Exactly the same equation is obtained
for the vector potential $\Aa(\rr,\omega)$, defined as $\nabla\times\Aa(\rr,\omega)=\mu(\rr)\HH(\rr,\omega)$,
if the Coulomb gauge $\nabla\cdot\Aa=0$ is applied \cite{viviescas03}.}
\ba\label{helm3}
\frac{\omega^2}{c_0^2}\EE(\rr,\omega)&=&M_\epsilon(\rr)\nabla\times [M_\mu(\rr)\nabla\times \EE(\rr,\omega)]\nonumber\\
&=:&\LL_\EE \EE(\rr,\omega)\, .
\ea
The operator $\LL_\EE$ on the RHS of this equation is not Hermitian, if
the (``naive'') definition of the scalar product
$<\EE_1|\EE_2>=\int d^3\rr \EE_1^*(\rr)\cdot\EE_2(\rr)$ is used. Only if we define
\cite{franosch20}
\be
<\EE_1|\EE_2>:=
\int d^3\rr 
M_\epsilon^{-1}(\rr)
\EE_1^*(\rr)\cdot\EE_2(\rr)\, ,
\ee
the operator $\LL_\EE$ has the Hermitian property:
\ba\label{combined}
&&<\EE_1|\LL_{\EE}\EE_2>\,\,=
\int d^3\rr\,
\EE_1^*(\rr)\cdot\big[
	\nabla\times M_\mu(\rr)[\nabla\times \EE_2(\rr)]\big]\nonumber\\
&&\quad =
\int d^3\rr\, M_\mu(\rr)
\big[\nabla\times \EE_1^*(\rr)\big]\cdot\big[\nabla\times \EE_2(\rr)\big]\nonumber\\
&&\quad =
\int d^3\rr\,
\EE_2(\rr)\cdot\big[
	\nabla\times M_\mu(\rr)[\nabla\times \EE_1^*(\rr)]\big]\nonumber\\
&&\quad\stackrel{!}{=}<\EE_2|\LL_{\EE}\EE_1>^{\!*}\, .
\ea
The second line guarantees the positiveness of the spectrum.
It is easily verified that for the scalar product without the
fluctuating permittivity included, $\LL_\EE$ is not Hermitian, because
extra terms involving $\nabla M_\epsilon$ are obtained.

Similarly an equation for the magnetic field can be derived
from Maxwell's equations
\ba\label{helm4}
\frac{\omega^2}{c_0^2}\HH(\rr,\omega)&=&M_\mu(\rr)\nabla\times M_\epsilon(\rr)[\nabla\times \HH(\rr,\omega)]\nonumber\\
&=:&\LL_\HH\HH(\rr,\omega)\, .
\ea
Here, the operator $\LL_\HH$ is Hermitian, if the scalar product
includes a factor $M_\mu^{-1}(\rr)$. In the case of pure electric
disorder ($M_\mu=const$) no special definition of the scalar product
is needed. This (properly defined)
eigenvalue equation for electric disorder was used recently for
treating transverse two-dimensional AL
\cite{schirm18}.

It is remarkable \cite{viviescas03} that for $\omega\neq 0$ Eqs. (\ref{helm3})
and (\ref{helm4}) automatically guarantee the transversality conditions
\be
\nabla\cdot\big[\EE(\rr,\omega)/M_\epsilon(\rr)\big]=0;\quad
\nabla\cdot\big[\HH(\rr,\omega)/M_\mu(\rr)\big]=0\, .
\ee

In order to formulate an analytic theory for the disorder-averaged physical
quantities in a system described by (\ref{helm3}) and (\ref{helm4})
it is rather disadvantageous to work with the disorder dependent
scalar product. This can be avoided
using symmetrized fields \cite{dyson53,viviescas03,rotter17}
$\Er := \EE /\sqrt{M_\epsilon(\rr)}$ and
$\Hr := \HH /\sqrt{M_\mu(\rr)}$ 
which obey the symmetrized Helmholtz equations
\ba\label{helm5}
\frac{\omega^2}{c_0^2}\Er(\rr,\omega)&=&
M_\epsilon^{1/2}(\rr)\nabla\times M_\mu(\rr)[\nabla\times M_\epsilon^{1/2}(\rr)
\Er(\rr,\omega)]
\nonumber\\
&=:&\LL_{\Er} 
\Er(\rr,\omega)\, ,
\ea
\ba\label{helm6}
\frac{\omega^2}{c_0^2}\Hr(\rr,\omega)&=&
M_\mu^{1/2}(\rr)\nabla\times M_\epsilon(\rr)[\nabla\times M_\mu^{1/2}(\rr)
\Hr(\rr,\omega)]
\nonumber\\
&=:&\LL_{\Hr}
\Hr(\rr,\omega)\, .
\ea
Eqs. (\ref{helm5}) and (\ref{helm6}) now constitute conventional
eigenvalue equations with operators $\LL_{\Er},\LL_{\Hr}$ that are
Hermitian with respect to the scalar products 
$<\Er_1|\Er_2>\,=\int d^3\rr \Er_1^*(\rr)\cdot\Er_2(\rr)$ and 
$<\Hr_1|\Hr_2>\,=\int d^3\rr \Hr_1^*(\rr)\cdot\Hr_2(\rr)$.

In this transformed way the differential operators
are manifestly Hermitian with respect to the conventional
definition of the scalar product. In this form the eigenvalue
problem
can be dealt with in the usual way, using
functional integrals and replica theory \cite{mckane81,kohler13}.

Generalizing the derivation of K\"ohler {\it et al.} \cite{kohler13}
we establish a coherent-potential approximation
(CPA), based on Eqs. 
(\ref{helm5}),
(\ref{helm6}),
along the lines of our peviou s work on elasticity.

The CPA arises as a saddle-point equation of an effective
field theory, constructed by field-theoretic methods \cite{kohler13}.
This variational derivation is equivalent to the traditional
method \cite{elliott74}
requiring that the scattering $T$ matrix of the
``perturbation'' $M_{\alpha,i}-M_{\alpha}(z)$, ($\alpha=\epsilon,\mu$)
be zero on the average.
In the CPA
the disordered system is replaced by an effective medium,
in which the fluctuating quantities (in our case
$M_\epsilon(\rr)$ and $M_\mu(\rr)$)
are replaced by uniform, but frequency-dependent, complex quantities
$M_\epsilon(z)$ and $M_\mu(z)$, where $z=\frac{1}{c_0}\omega+i\eta$, ($\eta$
is an infinitesimal positive real number),
except inside a cavity around the midpoint  
$\rr_i$. The volume of the cavity is 
$V_c$, and in this region
$M_{\epsilon,\mu}$ take their fluctuating values evaluated at $\rr_i$
$M_{\epsilon,i} \doteq M_{\epsilon}(\rr_i)$
and $M_{\mu,i} \doteq M_{\mu}(\rr_i)$. 
Within CPA these quantities are assumed to be uncorrelated\footnote{%
	\new{A
generalization of the traditional CPA for electrons 
for the inclusion of
{\it correlated} disorder exists \cite{zimmermann09}.
In this treatment the $k$ integral in Eq. (\ref{green1}) up to
the cutoff $k_\xi\propto \xi^{-1}$ has to be replaced by an integral
over the 
Green's function $G(k,z)$, multiplied by
the $k$ dependent correlation function $C(k)$, normalized
by its value at $k$ = 0. This function equals 1
for wavenumbers $k\ll k_\xi$ and then smoothly decays near
$k = k_\xi$. So, the present CPA just replaces this ``smooth cutoff''
of the correlated treatment
by a sharp one. The
essential ingredient of the spatial
correlations, namely the correlation length is included
in the present version of the CPA.
Long-range correlations, which are
relevant in hyperuniform
materials 
\cite{haberko20,scheffold22,montsarrat21},
and which govern the $k\rightarrow 0$ behavior
of the correlations,
are not included.
}},
which means that $V_c$ must be larger than the correlation
volume $\xi^3$, where $\xi$ is the correlation length. This. naturally introduces
an ultraviolet wavenumber cutoff $k_\xi\propto\xi^{-1}$ 
into the effective medium.
\new{In our treatment, this cutoff replaces the radius of the first Brillouin zone (in crystals) and the Debye cutoff (in glasses) for the definition of the
density of states $g(\omega)$
which samples the states relevant for the disorder scattering:}
\be
g(\omega)=
2\omega\rho(\lambda)=2\omega\frac{1}{\pi}\im\big\{
	G(z)\big\}\, ,
\ee
\new{where $\rho(\lambda)$ is the density of levels (eigenvalues), 
$G(z)$ is the local Green's function}
\be
G(z)=\label{green1}
\frac{3}{k_\xi^3}
	\int_0^{k_\xi}dk k^2
G(k,z)\, ,
\ee
\new{and
$G(k,z)$ is the wavenumber dependent Green's function of the
effective medium}
\be\label{green2}
G(k,z)=\frac{1}{-z^2+k^2M_\varepsilon(z)M_\mu(z)}\, .
\ee

We emphasize that -- in contrast to the PT treatment using the
nonlinear-sigma-model theory \cite{john84,john87} - 
\new{in CPA} the small parameter
for justifying the saddle-point approximation is not the relative
variance of the fluctuating quantities \cite{mckane81}, but
the ratio $V_c/V$ between the cavity volume and the
volume $V$ of the sample
\cite{kohler13}. This enables to treat the case of strong disorder, where
the relative variance may take any value.

The CPA equations read \cite{kohler13}

\be\label{cpa1}
0=\left\langle\frac{M_{\epsilon,i}-M_\epsilon(z)}{1+q\big(M_{\epsilon,i}-M_\epsilon(z)\big)\Lambda_\epsilon(z)}\right\rangle_\epsilon
\ee
and
\be\label{cpa2}
0=\left\langle\frac{M_{\mu,i}-M_\mu(z)}{1+q\big(M_{\mu,i}-M_\mu(z)\big)\Lambda_\mu(z)}\right\rangle_\mu
\ee
with $q=V_ck_\xi^3/3\pi^2$. 
The parameter $q$ must be smaller than 1 and
can be interpreted as a mean-field critical percolation
threshold \cite{kohler13}. Because the
critical percolation threshold for 3-dimensional continuum percolation
is around 0.3, we take $q=0.3$ in the numerical calculations that
we performed to show graphically the effect of the disorder.

The quantities $\Lambda_{\epsilon,\mu}(z)$ are defined by
\be
\Lambda_{\epsilon,\mu}(z)=\frac{1}{M_{\epsilon,\mu}(z)}
\big[1+z^2G(z)
	\big]
\ee

We note that the CPA 
equations (\ref{cpa1}) and (\ref{cpa2})
are completely symmetric with respect to
$\epsilon$ and $\mu$, i.e. they hold for both, Eqs
(\ref{helm5}) and
(\ref{helm6}). 
We further note that if the distributions of the two spatially fluctuating quantities are the same,
$\PP(M_{\epsilon,i})$ =
$\PP(M_{\mu,i})$, it results $M_\epsilon(z)=M_\mu(z)$. Therefore the CPA
equations reduce to the ones one would obtain if one
would take $M_\epsilon(\rr)=M_\mu(\rr)$ from the outset \cite{schirm22a}.

The averages $\langle\dots\rangle_{\epsilon,\mu}$ are to be performed
with distribution densities
$
\PP_{\epsilon}(M_{\epsilon,i})
$
and
$
\PP_{\mu}(M_{\mu,i})
$.
For our calculations, in order to be able to treat the case
of strong disorder, we take log-normal distributions \cite{kohler13}
$\PP(x)=[\sqrt{2\pi}\sigma x]^{-1}e^{-\ln^2(x/x^{(0)})/2\sigma^2}$
with medians 
\mbox{$x^{(0)}=M_\epsilon^{(0)}\!=\!
M_\mu^{(0)}\!=1$.}
The relative variances of the two distributions
$\gamma_\epsilon=
\langle(M_\epsilon-\langle M_\epsilon\rangle)^2\rangle/\langle M_\epsilon\rangle^2=
e^{\sigma_\epsilon^2}-1$ and 
$\gamma_\mu=
\langle(M_\mu-\langle M_\mu\rangle)^2\rangle/\langle M_\mu\rangle^2=
e^{\sigma_\mu^2}-1$
are the control parameters of the theory.

From the Green's function (\ref{green2})
we can read off the formula for the (scattering) mean-free path
\be
\frac{1}{\ell(\omega)}=\frac{2\omega}{c_0}\im\left\{
	\frac{1}{\big[M_\epsilon(z)M_\mu(z)\big]^{1/2}}
	\right\}
\ee
and the speed of light inside the medium:
\be
v(\omega)=c_0 \; \re\big\{\big[M_\epsilon(z)M_\mu(z)\big]^{1/2}
	\big\}
\ee
In turn, from these quantities we can calculate the frequency-dependent
(unrenormalized)
diffusivity
\be
D_0(\omega)=\frac{1}{3}v(\omega)\ell(\omega)\, .
\ee

\begin{figure}
	\includegraphics[width=8cm]{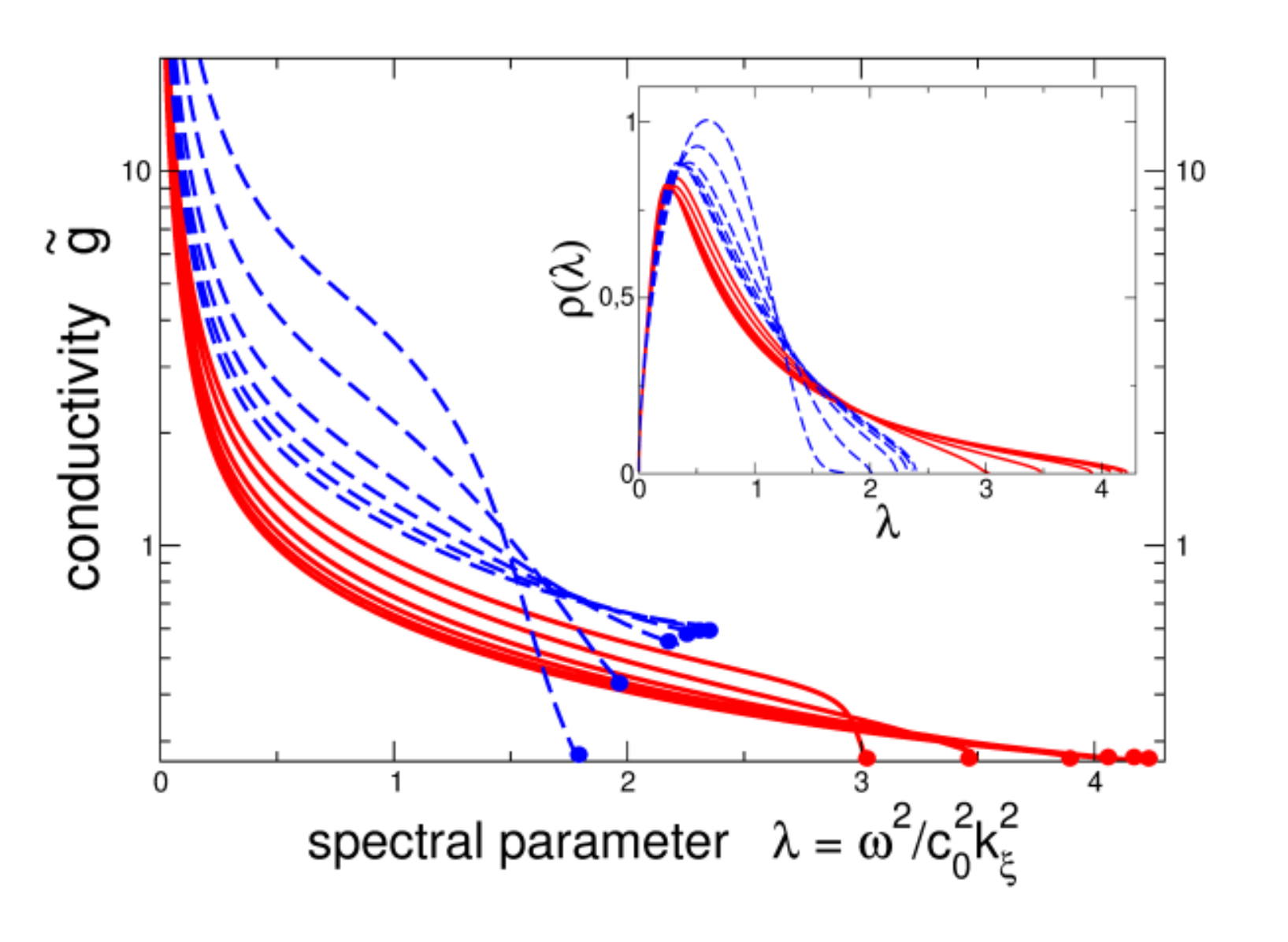}
	\caption{``Conductivity'' $\tilde g(\omega)=g(\omega)D(\omega)$
	against frequency, calculated in CPA for a log-normal distribution
	of $M_\epsilon$ and $M_\mu$,
	truncated at $M_\epsilon=M_\mu=0$. Dashed blue lines: Only
	one quantity, say $M_\epsilon(\rr)$ is fluctuating, the relative
	variance $\gamma_\epsilon=e^{\sigma_\epsilon^2}-1$ increases as
	$\gamma_\epsilon$ = 0.25, 0.5, 1., 1.5, 2. 2.5. \newline
	Continuous red lines: Both quantities $M_\epsilon(\rr)$ and $M_\mu(\rr)$
	are fluctuating, one of the variances, say, $\gamma_\epsilon$ is held
	fixed at 2.5, the other variance increases from $\gamma_\mu$ = 0.25
	to $\gamma_\mu$ = 2.5. in steps as before.\newline
	Inset: density of eigenvalues
	$\rho(\lambda)$ for the same CPA calculations.\newline
	The full circles in the main panel mark the end of the spectrum,
	given by $\rho(\lambda)$ in the inset.
	}
	\label{conductance}
\end{figure}

Before we use the CPA for estimating the localization properties
of disordered electromagnetic systems at finite frequency $\omega$,
we would like to comment
on the limit $\omega\rightarrow 0$. As pointed out by
K\"ohler {\it et al.}
\cite{kohler13}, in this limit the effective-medium expression
of Bruggeman \cite{bruggeman35} for the permittivity of mixed
dielectric materials is obtained. Contrary to this,
the CPA applied to the potential-type treatment of Maxwell's
equation \cite{sheng06}, mentioned in the
beginning, gives just the arithmetic
average of the permittivity in the $\omega\rightarrow 0$ limit,
because the non-trivial influence of the disorder in this
approach is multiplied by $\omega^2$ and just vanishes in
the DC limit. This shows once more that a proper treatment
of Maxwell's equations is necessary.

\new{We now turn to the discussion of the impact of electrical
and magnetic disorder on Anderson localization of light.
This phenomenon is known \cite{abrahams79} to arise from interference
of closed scattering paths.
According to the self-consistent theory of Anderson localization
\cite{vollhardt80,vollhardt82,wolfle10}
in the version used for classical waves
\cite{akkermans85,chu88,schirm93,montsarrat21}
the renormalized diffusion coefficient, which includes the localization
phenomena,
is given by
\be\label{self1}
D(\Omega,\omega)=D_0(\omega)-D(\Omega,\omega)P_0(\Omega,\omega)
\ee}
\new{Here $\Omega$ denotes the frequency corresponding to the
diffusion dynamics of the radiation, and $P_0(\Omega,\omega)$
denotes the return probability}
\new{\be
P_0(\Omega,\omega)=\frac{1}{\pi g(\omega)}
\sum_{|\qq|<q_0}
\frac{1}{-i\Omega+q^2D(\Omega,\omega)}\, .
\ee}
\new{The upper cutoff $q_0$ has been introduced, because the
interference is only effective in the $\qq$ region, where
the diffusion approximation holds. In the original
papers on electron localization
\cite{vollhardt80,vollhardt82,wolfle10}
the inverse mean-free path $\ell^{-1}$ has been taken
for $q_0$, in the literature on phonon localization
\cite{chu88,schirm93}
the Debye cutoff $k_D$, instead. Here we choose to take
the correlation cutoff $q_0=k_\xi$ as upper cutoff.
The self-consistent Eq. (\ref{self1}) can now be written in the form
\be\label{self2}
D(\Omega,\omega)=D_0(\omega)
-\frac{3}{\pi k_\xi^3g(\omega)}\int_0^{k_\xi}dq\frac{q^2}{q^2-\frac{i\Omega}{D(\Omega,\omega)}}
\ee}
\new{Localization or otherwise is now defined to occur if
the quantity
\be
\lim_{\Omega\rightarrow 0}D(\Omega,\omega)
\ee
vanishes or not.}

\new{We now assume that a frequency $\omega^*$ exists (mobility edge)
which separates the extended states ($\omega<\omega^*$) from the
localized ones ($\omega>\omega^*$). In the localized regime
the quantity $-i\Omega/D(\Omega,\omega)$ becomes a real quantity,
namely the square of the inverse localization length. Right at the
mobility edge $\omega=\omega^*$, this quantity becomes zero, and we have
\be\label{self3}
D(\Omega,\omega)=\bigg[
	1-\frac{3}{\pi k_\xi^2g(\omega)D_0(\omega)}
	\bigg]
\ee
On the other hand, at the mobility edge,
\mbox{$D(\Omega,\omega)=0$,} so that the 
dimensionless quantity (``conductivity'')
\be
\tilde g(\omega)=k_\xi^2g(\omega)D_0(\omega)
\ee
has to be equal to $3/\pi\approx 1$ at the mobility edge.}
Values of
$\tilde g(\omega)$ larger
than $\sim$1, therefore, lead to delocalization, 
values smaller than $\sim$1 to
localization.

In Fig. \ref{conductance}
we have plotted this quantity, calculated
in CPA against the dimensionless spectral
parameter 
$\lambda=\omega^2/c_0^2k_\xi^2$.
We consider
two scenarios:

\begin{itemize}
	\item[($i$)]
Only one of the moduli, say, $M_\epsilon(\rr)$
is considered to have spatial fluctuations with
variance 
$\gamma_\epsilon$ increasing from 0.25 to 2.5:
		{\it electric (or magnetic) disorder only}
		(dashed blue lines).
	\item[($ii$)]
Setting one of the variances, say, $\gamma_\mu$ = 2.5
and increasing the other, $\gamma_\epsilon$ from 0.25 to 2.5:
		{\it Combined electric and magnetic disorder}
		(continuous red lines).
\end{itemize}

It is seen that in the case of the combined electric and magnetic
disorder the values of $\widetilde g$ are much lower and also
the {\it spectral range} for which $\widetilde g$ is smaller than
$\sim 1$ is much more extended. Fig. \ref{conductance}
comprises the central result of
the present contribution.  Our results may explain, why with 
electric disorder only  (or perhaps magnetic disorder only)
it is very hard to obtain Anderson
localization, whereas for the combination of both, the odds for observing
Anderson localization of light in three dimension are increased
appreciably.

We therefore recommend for meeting the challenge of experimentally
observing 3D Anderson localization the consideration of
disordered materials with both electric and magnetic disorder.
Such materials could be e.g. polymeric materials with
superparamagnetic inclusions \cite{schulz10}.

Let us now discuss the recent numerical
results of Yamilov {\it et al.} \cite{yamilov22} in the light of our
findings. 
The authors considered two cases of
systems with the disorder induced by
overlapping spherical obstacles. These spheres were designed to have
in the first case
a high electric permittivity, in the second case
perfect electric conduction inside the spheres. 
In their first system with high dielectric permittivity of the spheres
they consider the case of
electric disorder only. In agreement with our results they
find no localization. On the other hand, by using perfectly conducting
obstacles they completely expel the time-varying electric and
magnetic
fields from
the obstacles, just effectively introducing a combination of electric
and magnetic disorder. Thus their numerical
observation of Anderson localization
for the perfectly conducting obstacles
corresponds to our prediction of localization for the
case of combined electric and magnetic disorder.

Summarizing, we have presented a mean-field theory for combined
electric and magnetic disorder based on eigenvalue equations
derived from Maxwell's equations, which involve
manifestly Hermitian operators. The results for the dimensionless
conductance suggest systems with combined electric and magnetic
disorder as candidates for 3D Anderson localization.

\section*{Acknowledgement}
GR is grateful to the European Research Council, Synergy Grant ASTRA  855923.
TF acknowledges support by the Austrian Science Fund (FWF): I 5257.

\end{document}